\documentstyle[multicol,aps,epsf,psfig]{revtex}
\begin{document}
\twocolumn[\hsize\textwidth\columnwidth\hsize\csname
 @twocolumnfalse\endcsname
\title{Unified Scaling Law for Earthquakes}
\author{Per Bak$^1$, Kim Christensen$^2$, Leon Danon$^2$, and Tim Scanlon$^2$}
\pagestyle{myheadings}
\address{{$1$}Department of Mathematics, Imperial College, Queen's Gate,
London SW7 2BZ, U.K. \\
{$^2$}Blackett Laboratory, Imperial College, Prince Consort Road,
London SW7 2BW, U.K.}

\maketitle
\date{\today}
\begin{abstract}
We show that the distribution of waiting times between earthquakes
occurring in California obeys a simple unified scaling law
valid from tens of seconds to tens of years, see Eq. (1) and Fig. 4.
The short time clustering,
commonly referred to as aftershocks, is nothing but the short time limit
of the general hierarchical properties of earthquakes. There is no
unique operational way of distinguishing between main shocks and
aftershocks. In the unified law, the Gutenberg-Richter $b$-value,
the exponent $-1$ of the Omori law for aftershocks, and the fractal
dimension $d_f$  of earthquakes appear as critical indices.
\end{abstract}
\vskip2pc]

Earthquakes are a complicated spatio-temporal phenomenon.
The number of earthquakes with a magnitude $M>m$
is given by the Gutenberg-Richter law \cite{2}.
In addition to the regularity in the rate of occurrence, 
earthquakes display a complex spatio-temporal behavior 
\cite{13,15}.
The spatial distribution of epicentres is fractal and they occur
on a fractal-like structure of faults \cite{15,3}.
Short-range temporal correlations between earthquakes are expressed
by Omori's law \cite{1},
which states that immediately following a main earthquake there is a
sequence of aftershocks whose frequency decays with time as $T^{-\alpha},
\alpha \approx 1$. This has led to the commonly held belief that
aftershocks are caused by a different relaxation mechanism than the
main shocks. 

The observed temporal complex behavior is obviously of dynamical origin.
However, the statistics of earthquakes as well as the geometrical fractal
structure displayed by the faults and by the spatial distribution of
epicenters is also a result of a dynamical process
and one might speculate whether it is possible to unify these
observations.

We propose a unified scaling law for the waiting times between earthquakes,
expressing a hierarchical organization in time, space, and magnitude.
There is a correlated regime where the distribution of waiting times
between earthquakes is a power-law, $T^{-\alpha}, \alpha \approx 1$
and an uncorrelated regime. However, the waiting time interval for
the crossover between the two regimes for earthquakes larger than a
given magnitude depends on the area and magnitude under consideration.

\begin{figure}
\vspace*{-1.75cm}
\centerline{\psfig{figure=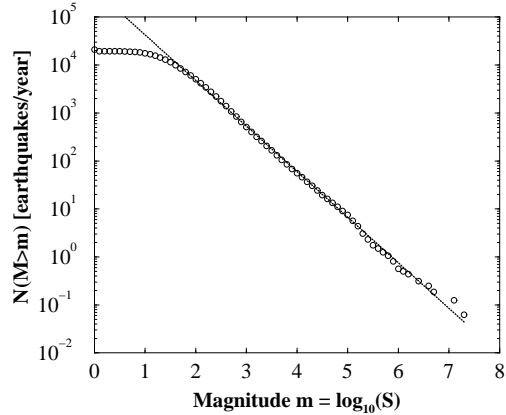,height=7cm,angle=-90}}
\caption{The number of earthquakes $N(M>m)$ with a magnitude larger than $m$
per year (open circles). The dashed line is the Gutenberg-Richter law
${\log}_{10} N(M>m) \propto -b m, b = 0.95$. The deficit at small
magnitude $m \leq 2$, is related to the problems with detecting small
earthquakes, so only earthquakes with $m \geq 2$ will be considered.
}
\label{fig1}
\end{figure}

\begin{figure}
\vspace*{-1.75cm}
\centerline{\psfig{figure=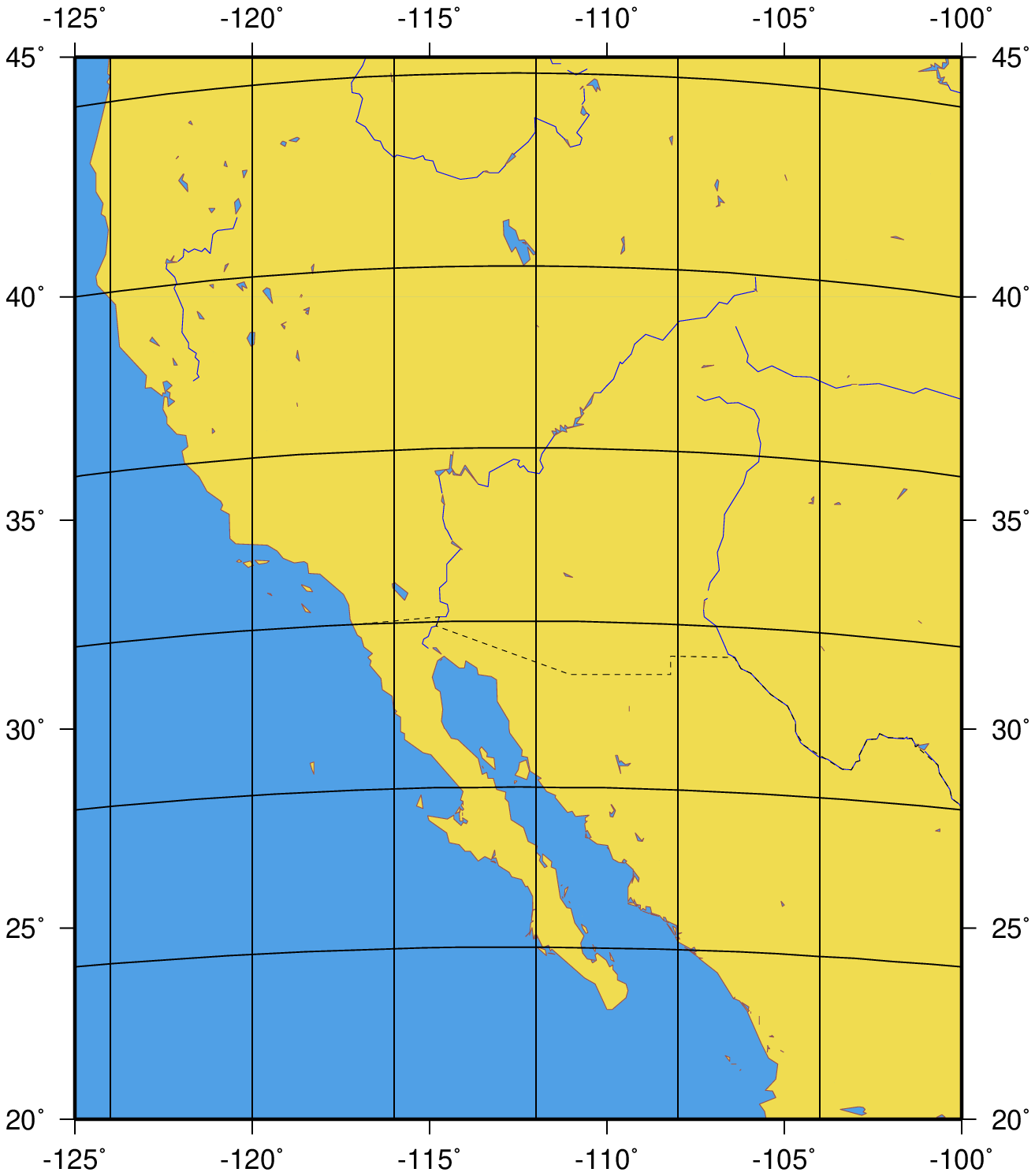,height=7cm,angle=0}
\hspace*{0.5cm}{\psfig{figure=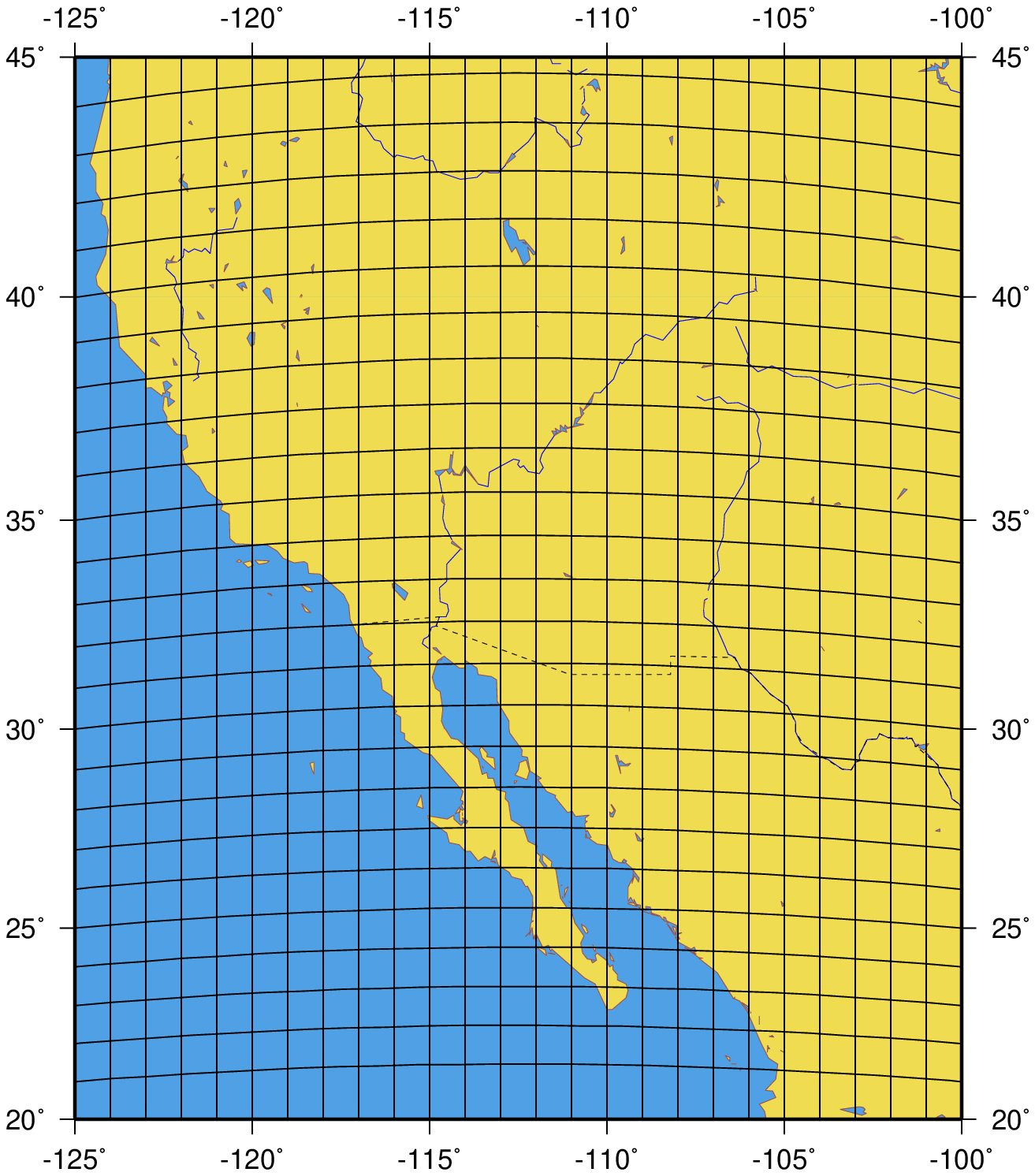,height=7cm,angle=0}}}
\caption{The region of California divided into grids of two
 different cell sizes (a) $L = 4^{\circ}$ and (b) $L = 1^{\circ}$.
}
\label{fig2}
\end{figure}

\begin{figure}
\vspace*{-1.75cm}
\centerline{\psfig{figure=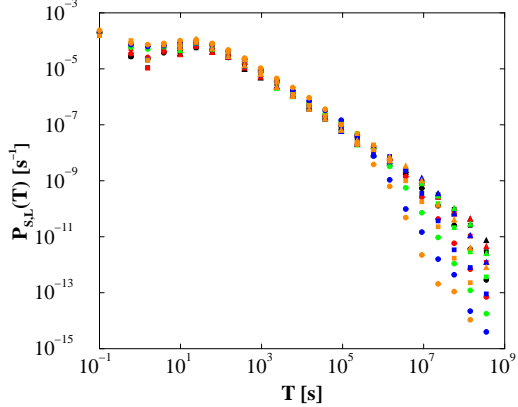,height=7cm,angle=-90}}
\caption{
The distribution $P_{S,L}(T)$ of inter-occurrence time
$T$ between earthquakes with magnitude $m$ greater than $log_{10}(S)$
within an area of linear size $L$. The solid circles, squares, and triangles
correspond to cut-offs $m = {\log}_{10}(S) = 2, 3$ and $4$, respectively.
The color coding represents the linear size $L = 0.25^{\circ}$ (black),
$0.5^{\circ}$ (red), $1^{\circ}$ (green), $2^{\circ}$ (blue),
and $4^{\circ}$ (orange) of the cells covering California.
For $T < 40$ s, earthquakes overlap and individual earthquakes cannot
be resolved. This causes the deficit for small $T$, so only intervals
$T > 38$ s will be considered in the following.
Notice, that for fixed cut-off magnitude $m$ but decreasing linear size $L$,
the deviation from the Omori law
$T^{-\alpha}, \alpha \approx 1$ sets in
for larger values of $T$. On the contrary, with fixed linear
size $L$ but decreasing cut-off magnitude $m$, the deviation from 
the Omori law $T^{-\alpha}, \alpha \approx 1$ sets in for smaller
values of $T$.
}
\label{fig3}
\end{figure}

\begin{figure}
\vspace*{-1.75cm}
\centerline{\psfig{figure=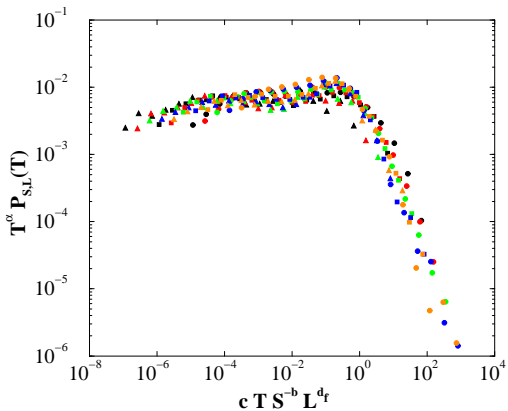,height=7cm,angle=-90}}
\caption{
The data in Fig. \ref{fig3} with $T > 38$ s re-plotted with
$T^{\alpha} P_{S,L}(T)$ as a function of the variable
$x = c\, T S^{-b} L^{d_f},  c = 10^{-4}$. The data-collapse implies
a unified law for earthquakes. The Omori law exponent
$\alpha = 1$, Gutenberg-Richter value $b = 1$, and fractal
dimension $d_f = 1.2$ have been used in order to collapse all
the data onto a single, unique curve $f(x)$. The estimated uncertainty in
the exponents is less than $0.2$. The function $f$ is constant
for $x < 1$, corresponding to the correlated Omori law regime,
while it is decaying fast for large arguments $x > 1$, associated with
the uncorrelated regime of earthquakes. Whether two earthquakes are
to be categorized as belonging to a correlated or uncorrelated sequence
not depend independently on the values of $T, S$ and $L$ but only on
the value of the product $x = T S^{-b} L^{d_f}$.
}
\label{fig4}
\end{figure}

An earthquake catalogue covering the period $1984 - 2000$ in a region
of California spanning 20$^{\circ}$N - 45$^{\circ}$N latitude
and 100$^{\circ}$W - 125$^{\circ}$W longitude was analyzed \cite{4}.
The total number of recorded earthquakes in the catalogue is $335,076$.
The number of earthquakes $N(M>m)$ with magnitude larger than $m$ is
given by the Gutenberg-Richter law \cite{2}
${\log}_{10} N(M>m) \propto -b m$, $b \approx 0.95$ see Fig. \ref{fig1}.
The spatio-temporal analysis was carried out as follows.
We covered the region with a grid with cells of size $L \times L$,
see Fig. \ref{fig2} and defined the waiting time $T$
as the time interval between the beginning of two successive earthquakes.
We then measured $P_{S,L}(T)$, the distribution of waiting times $T$,
between earthquakes occurring within range L whose magnitudes
are greater than $m = \log(S)$ .
\twocolumn[\hsize\textwidth\columnwidth\hsize\csname
 @twocolumnfalse\endcsname
\vskip2pc]
Figure \ref{fig3} shows the resulting set of curves
$P_{S,L}(T)$, for time scales ranging from seconds to
$16$ years, for several values of $S$ and $L$, plotted
on double logarithmic scale.
Obviously, the curves differ widely.
Some general trends can be seen, however. There is a linear regime,
indicating a power-law distribution, extending up to a cut-off indicating
an upper limit of the waiting time. For fixed cell size $L$ and
increasing cut-off $S$ (or $m$), the range of the power-law regime increases.
For fixed cut-off $S$ and increasing cell size $L$, the range of the
power-law regime decreases \cite{5}.
In Fig. \ref{fig4}, the curves are re-plotted in terms of re-scaled
coordinates. The $x$-axis is chosen as $x = T S^{-b} L^{d_f}$, and the
$y$-axis represents $y = T^{\alpha} P_{S,L}(T)$. The rescaling causes
a shift of the curves in Fig. \ref{fig3} that depends on $L$ and $S$.
For suitable choice of the interval exponent $\alpha$, the magnitude
exponent $b$, and the spatial dimension $d_f$, all the data collapse
nicely onto a single well-defined curve $f(x)$, that is,
\begin{equation}
T^{\alpha} P_{S,L}(T) = f(T S^{-b} L^{d_f}).
\end{equation}

This equation expresses the unified scaling law for earthquakes.
The function $f(x)$ consists of a constant part and a decaying part,
separated by a sharp kink. The constant part corresponds to the linear,
power-law part in Fig. \ref{fig3} since we have multiplied
$P_{S,L}(T)$ with $T^{\alpha}$. Any deviation from power-law behavior
would show up dramatically in this type of plot. Nevertheless, the function
is approximately constant over $8$ orders of magnitude! The rapidly
decaying part is consistent with an exponential decaying function
implying an uncorrelated regime for large values of
$x$. This is indeed what one would expect on physical ground: earthquakes that
are separated by large enough distances or long waiting times will
be uncorrelated.

The index $\alpha \approx 1$ can be
identified as the Omori-law exponent for aftershocks,
$b \approx 1$ is the $b$-value in the Gutenberg-Richter law,
and $d_f \approx  1.2$ describes the $2d$ fractal
dimension of the location of epi-centers projected onto
the surface of the Earth.

The data collapse implies that the waiting time distribution depends
on $T, S$, and $L$ only through the variable $x$.
Only critical processes exhibit this type of data collapse,
known as scaling in critical phenomena \cite{6}, so our analysis
demonstrates that earthquakes are a self-organized critical (SOC)
phenomenon \cite{footnote,7,8}, as had been anticipated from the existence
of the Gutenberg-Richter law \cite{9,10,11,12}.
The data collapse shows that there is no separate
relaxation mechanism for aftershocks.
The three exponents $\alpha, b$, and $d_f$ characterizing earthquakes
emerge as critical indices in the unified law.
The estimated uncertainty in the values of the critical indices is less
than 0.2. Whether the critical
exponents vary with region and maybe even with time is an interesting
question that is outside the scope of the present letter but we urge
further studies in that direction.

Depending on the value of scaling argument $x$, and thus the chosen values of
$L$ and $m$ (or $S$), two successive earthquakes
will either be correlated, for $x$ small, (i.e. to the left of the kink in
Fig. \ref{fig4}),
or uncorrelated, for $x$ large (i.e. to the right of the kink in
Fig. \ref{fig4}).

Depending on the length scale $L$ of observations, and the magnitude
magnitude $m$ (or $S$) chosen, the correlated Omori $T^{-\alpha}$ regime
may range from seconds to tens of years (and probably much longer if
data were available). If the earthquakes are correlated they may be
interpreted as belonging to an aftershock sequence. If they are uncorrelated,
they may be interpreted as main events. However, this depends on $L$ and
$m$ through the variable $x = T S^{-b} L^{d_f}$ and has no absolute meaning.
Therefore, there is no unique way of characterizing earthquakes as aftershocks
or main events.

To summarize, the short time correlations given by Omori's law is just the
short time limit of a general hierarchical scaling phenomenon occurring at
all accessible time scales. Amazingly, the statistics of aftershocks
occurring within minutes of an earthquake can be simply related to the
statistics of earthquakes separated by tens of years!

One may think of the value of $m$ (or $S$) at the kink as being
a 'characteristic' magnitude of earthquakes. However, since the
position of the kink is a function
of $x$, not $m$ (or $S$), one cannot specify the magnitude of the
'characteristic' earthquake without at the same time specifying a time
scale and an area. However, there is no special time scale that can play
any absolute role for the dynamics of earthquakes, limited at the upper
end by the time scale of tectonic plate motion, and at the lower end by
the duration of earthquakes. 

How should one physically understand the fundamental law, Eq. (1)?
Let us first discuss the meaning of the scaling variable
$x = T S^{-b} L^{d_f}$. The quantity $S^{-b} L^{d_f}$ appearing in it's
definition is a measure of the average number of earthquakes per time
unit with magnitude greater than $m = log_{10}(S)$ occurring within
the range $L$.
Thus, $x$ is a measure of the average number of such earthquakes occurring
within a time interval $T$. The law states that the distribution of
waiting times depends only on this number.
When this number exceeds a well-defined value (the position of the kink
in Fig. \ref{fig4}), the earthquakes sharply become less correlated.

Think of earthquakes being generated by 'processes', each producing a
sequence of correlated earthquakes with a $T^{-\alpha}$ distribution.
These processes correspond to a sequence of avalanches in
self-organized critical models of complex phenomena. Visually, one might
think of the processes as the activity associated with dynamically changing
fault segment patterns. The law indicates that the crust operates in the
true SOC slow-driving regime \cite{14} where the individual processes
(avalanches) do not overlap.
Because of the non-zero driving rate, several spatially separated processes
are active simultaneously. The kink on the $f$-curve indicates the point
where one crosses into the regime where spatially independent earthquakes,
belonging to different processes, are sampled within a window spanned by
$T$ and $L$. For small enough $L$ and $T$ only a single, correlated
process is sampled.

{\em Acknowledgements.} We would like to thank C. Scholz for advising
us on earthquake catalogues. P.B. acknowledge helpful discussions with
M. Paczuski. Correspondence and requests for materials should be
addressed to P.B. (email: bak@alf.nbi.dk) or K.C.
(e-mail: k.christensen@ic.ac.uk).

\end{document}